\documentclass[epj]{webofc}
\usepackage[varg]{txfonts}   
\usepackage{dutchcal}
\usepackage{amsfonts}
\usepackage{amsmath}
\usepackage{subfigure}
\usepackage{graphicx}
\usepackage{hyperref}

\begin{document}
\title{Conformal Ward Identities and the Coupling of QED and QCD to Gravity}
%
%

\author{\firstname{Claudio} \lastname{Corian\`o}\inst{1}\fnsep\thanks{\email{claudio.coriano@le.infn.it}}, \firstname{Matteo Maria} \lastname{Maglio}\inst{1}\fnsep\thanks{\email{matteomaria.maglio@le.infn.it}}}

\institute{Dipartimento di Matematica e Fisica "Ennio De Giorgi", \\ Universit\`a del Salento and INFN-Lecce, \\ Via Arnesano, 73100 Lecce, Italy}

\abstract{We present a general study of 3-point functions of conformal field theory (CFT) in momentum space, following a reconstruction method for tensor correlators, based on the solution of the conformal Ward identities (CWIs), introduced in recent works. We investigate and detail the structure of the CWIs and their non-perturbative solutions, and compare them to perturbation theory, taking QED and QCD as examples. Exact solutions of CFT's in the flat background limit in momentum space are matched by the perturbative realizations in free field theories, 
showing that the origin the conformal anomaly is related to efffective scalar interactions, generated by the renormalization of the longitudinal components of the corresponding operators. } 

\maketitle
\section{Introduction}
Correlation functions of 3-point functions play a special role in conformal field theory (CFT) in $d=4$ since they can be determined by solving the  constraints imposed by the conformal group, which in $d=4$ is $ \text{SO}(2,4)$. The functional form of the solutions are determined more easily using methods of coordinate space rather than momentum space. Operator product expansion techniques, for instance, which allow to define a program for the bootstrap of higher point functions, as well as other analysis, have so far favoured coordinate space.\\
On the other hand, studies of CFT's based on the standard perturbative expansion in the gauge coupling, such as in $N=4$ supersymmetric Yang-Mills theory, as far as the computation of the amplitudes is concerned, have been largely performed in momentum space. A Lagrangian realization of a CFT is, in any case, a theory which is often investigated perturbatively in terms of its Feynman diagrams. On the contrary, a direct solution of the conformal Ward identities - for a given correlation function - can be interpreted, in this respect, as being strictly non perturbative, in the sense that it is not related to any coupling but only to the underlying symmetry.\\
Establishing a link between the two approaches in momentum space, with exact CFT solutions on one side and Lagrangian realizations on the other, turns useful for various reasons. One of them is to provide a simplification of general CFT results, as we are going to show next. A second one is the possibility of offering a simple and physical interpretation of the origin of the conformal anomaly in 3-point functions, for correlators with one or more insertions of stress energy tensors. 
\section{The origin of the conformal anomaly} 
Anomalies are associated, in coordinate space, to configurations of a correlator in which all the points coalesce. For a 3-point function these are configurations of the form $\sim \delta(x_1-x_2)\delta(x_2-x_3)$. Configurations of partial coalescence are defined to be {\em semilocal} or simply {\em contact} terms. Two counterterms are needed in $d=4$ in the renormalization of of correlators with multiple stress energy tensors, responsible for the generation of the scheme independent part of the anomaly: $E$ and $C^2$. They correspond to the Euler-Poincare density and the Weyl tensor 
squared respectively, which find application in the $TTT$ (3-graviton) vertex. For other correlators, such as the $TJJ$ vertex the renormalization 
of the vector 2-point function $JJ$ is sufficient to generate a finite correlator, which is reflected in the $F^2$ term of the anomaly 
functional, with $F$ the field strength of the photon (see \cite{Bastianelli:2012bz, Bastianelli:2012es} for related studies).
The solutions of the CWIs for 3-point functions in coordinate space have been obtained in the past by superposition of a traceless contribution and of contact terms proportional to the two operators mentioned above. 
Obviously, the coordinate space approach provides limited information on the origin of the anomaly, except for telling us that its origin is a short distance effect.

\section{Anomalies as light cone processes}
Once a CFT is matched to a free field theory, one has the possibility of handling  far more simplified expressions of such correlators and can proceed with the analysis of an ordinary Feynman expansion. As discussed in several previous works, the analysis in momentum space provides additional information on this phenomenon. This is described by the emergence of massless effective scalar degrees of freedom in 3-point functions containing insertions of stress energy tensors, which can be interpreted as light-cone interactions. 
As discussed in  \cite{Giannotti:2008cv,Armillis:2009pq,Armillis:2010qk}, this phenomenon clearly points towards an interpretation of the origin of the conformal anomaly as mediated by correlated pairs of fermions/scalars, as emerging from the spectral representation of a given perturbative correlator. More recently, it has been pointed out that such type of interactions play a role in the context of Weyl semimetals, with the paired electrons (representing the massless pole) interacting with the lattice of such materials \cite{Chernodub:2017jcp}.  Both chiral and conformal anomaly poles play a role in this phenomenon.\\ 
 Establishing conclusively the presence of such interactions at nonperturbative level is possible for 3-point functions, by matching the perturbative result with the nonperturbative solutions derived from the CWI's. Using this approach we have conclusively shown in \cite{Coriano:2018bbe} that these interactions are associated to renormalization and are not related to specific parameterization of the tensor correlators. The proof can be illustrated more easily in the case of the TJJ correlator in QED, that we will sketch below. A similar analysis can be performed for QCD, though it is more involved. 

\section{Effective interactions} 
The key construct in this analysis is the anomaly action, which for 3-point functions can be completely fixed, modulo few constants, which are specific to a given CFT. \\
It has been shown that in an uncontracted anomaly vertex of either chiral, conformal  \cite{Giannotti:2008cv,Armillis:2009pq,Armillis:2010qk} or superconformal type \cite{Coriano:2014gja}, the origin of an anomaly has to be attributed to the appearance
of specific form factors in its tensor structure, which are proportional to $1/k^2$
in the massless
limit. Such anomaly poles define massless exchanges in momentum space and are the direct
signature of the anomaly. In all such cases $k$ denotes the momentum of an axial-vector current
in an AV V (axial-vector/vector/vector) correlator or that of a stress energy tensor (T) in a
TJJ vertex.\\
The existence of chiral anomaly poles has been discussed in the literature in the 70's, while conformal anomaly poles have been shown to be part of the
TJJ vertex in QED, QCD and the electroweak sector of the Standard Model \cite{Armillis:2010qk,Coriano:2011zk} 
only more recently. 
Studies of such interactions in the context of both chiral and conformal anomaly diagrams
have always been performed at the perturbative level in the past, with the obvious limitations of
the case. These studies show the presence of some universal features of these interactions.
The chiral and conformal anomaly coefficients are then proportional to the residues of the corresponding
correlators evaluated at the anomaly pole (times a tensor structure which is the anomaly functional). By combining 
nonperturbative information coming from the solution of the CWI's with that from free field theories 
it is possible to show rigorously the existence of such interpolating states.  
\section{Conformal Ward identities in momentum space}
The CWIs in momentum space, in $d$ dimensions, for a correlation function, in the simpler case of a scalar correlator 
\begin{equation}
\Phi(x_1,x_2,\ldots x_n)=\langle \phi_1(x_1)\phi_2(x_2)\ldots \phi_3(x_n)\rangle 
\end{equation}
in momentum space take the form \cite{Coriano:2012wp,Bzowski:2013sza}
\begin{equation}
\label{sc1}
\left[\sum_{j=1}^n \Delta_j  -(n-1) d -\sum_{j=1}^{n-1}p_j^\alpha \frac{\partial}{\partial p_j^\alpha}\right]
\Phi(p_1,p_2,\ldots,\overline{p}_n)=0.
\end{equation}
(with $\bar{p}_n^\mu=-\sum_{j=1}^{n-1}p_j^\mu$)
for the dilatation and

\begin{equation}
\sum_{j=1}^{n-1}\left(p_j^\kappa \frac{\partial^2}{\partial p_j^\alpha\partial p_j^\alpha} + 2(\Delta_j- d)\frac{\partial}{\partial p_j^\kappa}-2 p_j^\alpha\frac{\partial^2}{\partial p_j^\kappa \partial p_j^\alpha}\right)\Phi(p_1,\ldots p_{n-1},\bar{p}_n)=0.
\end{equation}
for the special CWIs. We will choose $n=3$.
The latter equation can be rewritten in the form
\begin{equation}
K^{\kappa}_{scalar}\Phi\,(p_1,p_2, \bar{p}_3)=0, 
\end{equation}
in terms of the magnitudes of the momenta $ p_i\equiv \sqrt{p_i^2}$
with
\begin{equation}
{{K}}^{\kappa}_{scalar}=\sum_{i=1}^3 p_i^\kappa \,{K}_i 
\label{kappa2}
\end{equation}
and the scalar operator
\begin{equation}
{ K}_i\equiv \frac{\partial^2}{\partial    p_i \partial    p_i} 
+\frac{d + 1 - 2 \Delta_i}{   p_i}\frac{\partial}{\partial   p_i}.
\end{equation}
Eq. (\ref{kappa2})  can be split into the two independent equations
\begin{equation}
K_{13}\equiv K_2 - K_3=\frac{\partial^2\Phi}{\partial   p_1\partial   p_1}+
\frac{1}{  p_1}\frac{\partial\Phi}{\partial  p_1}(d+1-2 \Delta_1)-
\frac{\partial^2\Phi}{\partial   p_3\partial   p_3} -
\frac{1}{  p_3}\frac{\partial\Phi}{\partial  p_3}(d +1 -2 \Delta_3)=0.
\label{3k1}
\end{equation}
Following the approach presented in \cite{Coriano:2013jba}, the ansatz for the solution can be taken of the form 
\begin{equation}
\label{ans}
\Phi(p_1,p_2,p_3)=p_3^{\Delta - 2 d} x^{a}y^{b} F(x,y)
\end{equation}
with $x=\frac{p_1^2}{p_3^2}$ and $y=\frac{p_2^2}{p_3^2}$. Here we are taking $p_3$ as "pivot" in the expansion, but we could equivalently choose any of the 3  momentum invariants. $\Phi$ is required to be homogeneous of degree $\Delta-2 d$, where 
$\Delta=\Delta_1 +\Delta_2+\Delta_3$ is the total scaling dimension of $\Phi$. 
In the case of a scalar correlator the four fundamental solutions are expressed in terms of the Appell function 
\begin{equation}
\label{F4def}
F_4(\alpha(a,b), \beta(a,b); \gamma(a), \gamma'(b); x, y) = \sum_{i = 0}^{\infty}\sum_{j = 0}^{\infty} \frac{(\alpha(a,b))_{i+j} \, 
	(\beta(a,b))_{i+j}}{(\gamma(a))_i \, (\gamma'(b))_j} \frac{x^i}{i!} \frac{y^j}{j!} 
\end{equation}
where $(\alpha)_i = \Gamma(\alpha + i)/ \Gamma(\alpha)$ is the Pochammer symbol. The general solution is expressed as linear combinations of the 4 independent special solutions of type \eqref{ans} as \cite{Coriano:2012wp,Bzowski:2013sza}
\begin{equation}
\Phi(p_1,p_2,p_3)=p_3^{\Delta-2 d-2} \sum_{a,b} c(a,b,\vec{\Delta})\,x^a y^b \,F_4(\alpha(a,b), \beta(a,b); \gamma(a), \gamma'(b); x, y) 
\end{equation}
where the sum runs over the four values $a_i, b_i$ $i=0,1$ which define the Fuchsian exponents, with constants $c(a,b,\vec{\Delta})$, with $\vec{\Delta}=(\Delta_1,\Delta_2,\Delta_3)$. They have been shown in \cite{Coriano:2018zdo} to be identical for all the type of correlators investigated so far. The CWIs correspond to hypergeometric systems of equations which can be solved in two different ways \cite{Coriano:2018bbe,Bzowski:2013sza}.

\section{Tensor Correlators}
In the case of the $TJJ$ vertex, the proof in CFT that such poles are not artificial exploits two different representation of the same vertex. 
We sketch the proof.\\
In a basis of 13 form factors introduced in \cite{Giannotti:2008cv} the correlator takes the form
\begin{equation}
\Gamma^{\mu_1\nu_1\mu_2\mu_3}(p_2,p_3)\equiv \langle T^{\mu_1\nu_1}(p_1)\,J^{\mu_2}(p_2)\,J^{\mu_3}(p_3)\rangle=\sum_{i=1}^{13}\,F_i(s;s_1,s_2,0)\,t_i^{\mu_1\nu_1\mu_2\mu_3}(p_2,p_3),
\end{equation} 
in terms of a suitable set of independent tensor structures $t_i^{\mu_1\nu_1\mu_2\mu_3}$ and form factors $F_i$. The basis is non minimal, since it is not built by imposing the conformal constraints but only the conservation and symmetry properties of the correlator.
It contains only two form factors whose tensor structures are tracefull $(t_1,t_2)$ and only one form factor which needs to be renormalized as $d\to 4$, denoted as $F_{13}$. The form factors $F_i$ are functions of the kinematic invariants $s=p_1^2=(p_2+p_3)^2$, $s_1=p_2^2$, $s_2=p_3^2$. This representation of the $TJJ$ has been studied at 1-loop in QED, with an extension to QCD.\\
An alternative and minimal basis has been introduced in \cite{Bzowski:2013sza} using a decomposition of the same correlator in terms of transverse traceless ($tjj$) and longitudinal components ($t_{loc} JJ$) in the form
\begin{equation}
\langle T^{\mu_1\nu_1}(p_1)\,J^{\mu_2}(p_2)\,J^{\mu_3}(p_3)\rangle = \langle t^{\mu_1\nu_1}(p_1)\,j^{\mu_2}(p_2)\,j^{\mu_3}(p_3)\rangle +\langle t^{\mu_1\nu_1}_{loc}(p_1)\,J^{\mu_2}(p_2)\,J^{\mu_3}(p_3)\rangle
\end{equation}
with the transverse traceless sector expanded in a set of 4 form factors $A_l$ \cite{Coriano:2018bbe,Bzowski:2018fql, Bzowski:2015yxv}
\begin{align}
\langle t^{\mu_1\nu_1}(p_1)j^{\mu_2}(p_2)j^{\mu_3}(p_3)\rangle =&
{\Pi_1}^{\mu_1\nu_1}_{\alpha_1\beta_1}{\pi_2}^{\mu_2}_{\alpha_2}{\pi_3}^{\mu_3}_{\alpha_3}
\left( A_1\ p_2^{\alpha_1}p_2^{\beta_1}p_3^{\alpha_2}p_1^{\alpha_3} + 
A_2\ \delta^{\alpha_2\alpha_3} p_2^{\alpha_1}p_2^{\beta_1}\right. \notag\\
& \hspace{0.6cm}\left.  + 
A_3\ \delta^{\alpha_1\alpha_2}p_2^{\beta_1}p_1^{\alpha_3}+ 
A_3(p_2\leftrightarrow p_3)\delta^{\alpha_1\alpha_3}p_2^{\beta_1}p_3^{\alpha_2}
+ A_4\  \delta^{\alpha_1\alpha_3}\delta^{\alpha_2\beta_1}\right).\label{DecompTJJ}
\end{align}
The two basis can be related and the exact solution of the CWI's fixed up to 2 constants in $d=4$. The analysis of the exact solutions shows that the 
singularities in the $A_i$ basis are in exact correspondence with the presence of $F_{13}$, setting a direct link between the two basis. 
By imposing the traceless condition in $d$ dimensions one gets the constraints
\begin{equation}
\begin{split}
F_1&=\frac{(d-4)}{p_1^2(d-1)}\bigg[F_{13}-p_2^2\,F_3-p_3^2\,F_5-p_2\cdot p_3\, F_7\bigg]\\[1.2ex]
F_2&=\frac{(d-4)}{p_1^2(d-1)}\bigg[p_2^2\,F_4+p_3^2\,F_6+p_2\cdot p_3\,F_8\bigg]\,,
\end{split}\label{confeq}
\end{equation}
which as $d\to 4$ show that $F_2\to 0$ and $F_1\sim 1/p_1^2$ if $F_{13}$ develops a single pole in $1/\epsilon$, as is the case in 
a CFT. Indeed, in the $d\to 4$ limit 
\begin{equation}
\lim_{d\to 4} F_1=-\frac{4}{9}\frac{\pi^2}{p_1^2},
\end{equation}
showing the appearance of an anomaly pole in the single form factor which is responsible for the trace anomaly. It is then clear that the emergence of an anomaly pole in the $TJJ$ is not limited to perturbation theory but is a specific feature of the 
nonperturbative solution as well, if we can show that the general expression for the $TJJ$ can be exactly matched in free field theory \cite{Coriano:2018bbe,Coriano:2018zdo}. In the dispersive representation of the unique form factor which is responsible for the appearance of the anomaly, it is related to the exchange of a collinear fermion/antifermion pair in the $s$ variable $(\rho(s)\sim \delta(s))$, where $\rho(s)$ is the spectral density of the single form factor $F_1$ which contributes to the trace. The study can be performed in the non conformal limit, with a nonzero mass fermion in the loop, taking the massless limit at a second stage. This configuration provides a contribution to the anomaly action of the form 
\begin{equation}
 \label{pole}
\mathcal{S}_{pole}= - \frac{e^2}{ 36 \pi^2}\int d^4 x d^4 y \left(\square h(x) - \partial_\mu\partial_\nu h^{\mu\nu}(x)\right)  \square^{-1}_{x\, y} F_{\alpha\beta}(x)F^{\alpha\beta}(y)
\end{equation}
with the appearance of a massless nonlocal interaction.
In view of the equivalence between the perturbative and the nonperturbative solution for the $A_i$ (and henceforth for the $F$- basis), 
this behaviour is clearly present in the nonperturbative solution, due to the complete matching between the two \cite{Coriano:2018bbe,Coriano:2018zdo}.
\section{Conclusions} 
The breaking of conformal symmetry by the conformal anomaly is an aspect of QED and QCD which is of general significance. It is associated to effective massless interactions. This provides a new perspective on the analysis of the conformal phases of realistic field theories and their renormalization group flows in $D=4$. 


\end{document}